# The aqueous and crystalline forms of L-alanine zwitterion

Ivan Degtyarenko,*,[1] Karl J. Jalkanen,[2] Andrey A. Gurtovenko[3] and Risto M. Nieminen[1]

[1] Laboratory of Physics, Helsinki University of Technology, P.O.B. 1100, FIN-02015 HUT, Finland. Fax: +358 9 4513116; Tel: +358 9 4513110, +358 9 4513105; E-mail: imd@fyslab.hut.fi , rni@fyslab.hut.fi

[2] Nanochemistry Research Institute, Department of Applied Chemistry, Curtin University of Technology, G.P.O. Box U1987, Perth 6845, Western Australia. Fax: +61 8 9266 4699; Tel: +61 8 9266 7172; E-mail: jalkanen@ivec.org

[3] Computational Laboratory, Institute of Pharmaceutical Innovation, University of Bradford, Bradford, West Yorkshire, BD7 1DP, UK. Fax: +44 1274 234679; Tel: +44 1274 236114; E-mail: a.gurtovenko@bradford.ac.uk

**Abstract**

The structural properties of L-alanine amino acid in aqueous solution and in crystalline phase have been studied by means of density-functional electronic-structure and molecular dynamics simulations. The solvated zwitterionic structure of L-alanine ($^+NH_3$-$C_2H_4$-$COO^-$) was systematically compared to the structure of its zwitterionic crystalline analogue acquired from both computer simulations and experiments. It turns out that the structural properties of an alanine molecule in aqueous solution can differ significantly from those in crystalline phase, these differences being mainly attributed to hydrogen bonding interactions. In particular, we found that the largest difference between the two alanine forms can be seen for the orientation and bond lengths of the carboxylate ($COO^-$) group: in aqueous solution the C-O bond lengths appear to strongly correlate with the number of water molecules which form hydrogen bonds with the $COO^-$ group. Furthermore, the hydrogen bond lengths are shorter and the hydrogen bond angles are larger for L-alanine in water as compared to crystal. Overall, our findings strongly suggest that the generally accepted approach of extending the structural information acquired from crystallographic data to a L-alanine molecule in aqueous solution should be used with caution.

**1. Introduction**

L-alanine (LA) is the smallest, naturally occurring chiral amino acid with a non-reactive hydrophobic methyl group (-$CH_3$) as a side chain. LA has the zwitterionic form ($^+NH_3$-$C_2H_4$-$COO^-$) both in crystal and in aqueous solution over a large range of pH. The crystalline state is well defined structurally and can be successfully used for a detailed examination of a broad range of molecular properties. The crystal structure of alanine was studied by both X-ray (at T=298 K and T=23 K)[1,2] and neutron diffraction (at T=298 K).[3] The low-temperature data was used for a detailed analysis of the electrostatic properties of an alanine molecule.[4,5] X-ray diffraction,[6] infrared spectroscopy,[7] Raman scattering,[8] and coherent inelastic neutron scattering[9] were employed to study vibrational dynamics. The crystalline L- and D-alanine enantiomers were also used to investigate parity violation.[10,11]

In contrast to the success in studying an alanine in the solid state, experimental studies by infrared, Raman and neutron diffraction spectroscopic techniques have not yielded any conformational information about α-alanine in aqueous solution, having mainly focused on information concerning the structure of the hydration shell and the interactions between water molecules and amino acid residues.[12,13,14,15,16] Furthermore, the rotational (microwave) spectra in solution are not resolved to give structural information directly, so that the structural parameters of alanine molecules in water come predominantly from computational studies.[17,18,19]

As a result of this lack of experimental information regarding the alanine's structure in water, the experimental zwitterionic structure of alanine amino acid which is derived from solid state crystallographic data is often considered to be also valid for L-alanine in aqueous media. However, the origin for the L-alanine's zwitterionic form is different in crystal and in water. In crystals all three available protons of the ammonium group ($NH_3^+$) are used to form single N-H···O hydrogen bonds with oxygen atoms of three carboxylate groups ($COO^-$) of nearest amino acid molecules, thus linking the molecules together to form a three-dimensional crystal structure. In contrast, the key determinant of the stable zwitterionic structure in aqueous solution is hydrogen bonding interactions with surrounding water molecules (N-H···$O_w$ and O···$H_wO_w$ type of hydrogen bonds). The L-alanine zwitterionic species in water and in crystal are visualized in Figures 1 and 2.

[Figure 1 position]

[Figure 2 position]

To address the problem of a different nature of the alanine's zwitterionic forms in crystal and water, we have performed density-functional theory (DFT) electronic-structure and molecular dynamics calculations for L-alanine in aqueous solution and in crystalline phase. The effects of hydration are modelled explicitly by considering a relatively large number of water molecules (50) distributed around a L-alanine zwitterion. The dynamics of the whole system was treated fully quantum mechanically. We compared the calculated L-alanine zwitterion structures in water and in solid state. By means of analysis of molecular dynamics trajectories and charge density distribution we studied the role of intermolecular hydrogen bonding interactions and their influence on molecular properties.

## 2. Methods

### 2.1. Structures and Computational Methods

We have performed Born-Oppenheimer molecular dynamics (MD) simulations of a L-alanine amino acid in its zwitterionic form in aqueous environment. In order to stabilize the structure of the L-alanine zwitterion, the overall number of water molecules was chosen to be large enough to accommodate two solvation shells of the L-alanine molecule. Since *ab initio* modeling of an amino acid surrounded by a significant amount of water molecules is computationally very expensive, it is crucial to have a good initial guess regarding the number of water molecules, their positions and orientations around the amino acid. Therefore, we first performed classical molecular dynamics simulations of an L-alanine molecule solvated in a box with large amount of water molecules. Then we extracted a well-equilibrated structure of the L-alanine with a limited number of water molecules nearest to the amino acid; all other water molecules were removed. The resulting structure was then used for density-functional theory Born-Oppenheimer MD simulations. Overall, our simulations were performed in three steps.

(i) Classical atomic-scale MD simulations of a L-alanine in a box of water were first carried out with the use of an empirical force-field. Force-field parameters for the L-alanine were taken from the full-atom Gromacs force-field supplied within the GROMACS package (ffgmx2 set).[20] A molecule of L-alanine was solvated by around 500 water molecules; the simple point charge (SPC) model[21] was used to represent water. The MD simulations were performed in the N$p$T ensemble with temperature and pressure being kept constant with the use of the Berendsen scheme[22]. The temperature was set to 300 K and the pressure was set to 1 bar. The Lennard-Jones interactions were cut off at 1 nm. For the long-range electrostatic interactions we used the particle-mesh Ewald (PME) method.[23] The time step used was 1 fs and the total simulation time was 100 ps. The classical MD simulations were performed using the GROMACS suite.[20]

(ii) The final structure of the step (i) was used for preparing the initial structure for subsequent *ab initio* simulations. This was accomplished by removing water molecules located farther than 6 Å from any atom of the L-alanine. Using this criterion only 50 water molecules were eventually left around the L-alanine zwitterion. The resulting structure was then fully relaxed.

(iii) Finally, the first principle simulation were performed with the use of a numerical atomic orbitals DFT approach[24] as implemented in the SIESTA code.[25] The *ab initio* calculations were carried out within the generalized-gradient approximation, in particular, with the Perdew-Burke-Ernzerhof exchange-correlation functional (PBE)[26], and a basis set of numerical atomic orbitals at the double-z polarized level.[27] The choice of the exchange-correlation PBE functional used in our study is based on its reliability in the description of strong and moderate hydrogen bonds.[28] It is known to give accurate molecular bond lengths with the mean absolute error 0.012 - 0.014 Å.[29,30] Core electrons are replaced with norm-conserving pseudopotentials in their fully nonlocal representation.[31] The integrals of the self consistent terms of the Kohn-Sham Hamiltonian were obtained using a regular real space grid to which the electron density was projected. A kinetic-energy cutoff of 150 Ry was used for the MD run, which gives a spacing between grid points of ~ 0.13 Å. A cutoff of 300 Ry was used for structure optimization. The initial equilibration of the system was done at 300K; a thermostat was then switched off during production, so that the microcanonical ensemble was probed and flying ice cube effect is avoided.[32] The density-functional MD simulations were run for 40 ps with a time step of 1 fs; only the last 38 ps were used for the subsequent analysis. For visualization and trajectory analysis the Visual Molecular Dynamics (VMD) package was employed.[33] The same methodology has successfully been used in recent studies of problems of biological relevance, including modeling proteins, DNA and liquid water.[34]

The experimental non-optimized geometries from a neutron diffraction study at 298 K with precisely defined positions of hydrogen atoms have been adopted for all calculations with crystalline structures.[3] The crystal structures were obtained through the Cambridge Structural Database (CSD).[35] Unit cell constants ($a$ = 6.025, $b$ = 12.324, $c$ = 5.783 Å) were fixed and only atomic positions were relaxed.

## 3. Results and Discussion

*Crystalline and aqueous forms of L-alanine zwitterion.* The molecular geometry parameters determined from computer simulations are listed along with the experimental data in Table 1, and the atomic positions are shown in Figures 1 and 2. The experimental crystalline L-alanine bond lengths and angles, being corrected for libration, are based on neutron and X-ray diffraction values at T=298 K and T=23 K, respectively.[3,2] Values for simulated structures in water at room temperature were computed by averaging over molecular dynamics trajectories. In the analysis of trajectories the fact that atoms change their positions due to rotations was taken into account, so that the $O^1$ atom is always the one closer to the ammonium group and the $H^1$ atom is the one in between $CH^3$ and $CO^2$ groups.[36] For the calculated structures at T=0 K we used a conjugate gradient algorithm to define a minimum energy configuration.

[Table 1 position]

The noticeable difference between low-temperature experimental[2] and calculated crystalline structures is observed for the $O^1$-C'-$C_\alpha$-H, $H^5$-$C_\beta$-$C_\alpha$-H, and $H^3$-N-$C_\alpha$-$H_\alpha$ dihedral angles, see Figure 3. The corresponding carboxylate, methyl, and ammonium groups are rotated by 6.6, 4.5 and 2.5 degrees, respectively. The bond length changes range from 0.007 to 0.027 Å. Remarkably, only the $C_\alpha$-C', C'-$O^2$ and N-$H^{1,2,3}$ bonds lengthening turns out to significant, being 0.023, 0.02 and 0.027 Å, respectively. As demonstrated below, these structural changes correspond to hydrogen bonding optimization. The rest of the changes are well within the error bars typical of the method used 0.014 Å.[29] The absolute mean deviations for bond lengths and angles of the experimental and theoretical crystalline structures are 0.016 Å and 0.6 degree, respectively.[37]

The largest difference between the calculated optimized L-alanine structure in aqueous solution and the structure calculated in the crystalline phase is related to the carboxylate group. The overall absolute mean deviation between two phases at T=0 K are 0.017 Å and 1.4 degree (Figure 3).[37] The C'-O$^1$ bond lengthens by 0.04 Å and C'-O$^2$ shrinks by 0.023 Å. The overall temperature effect with respect to structural changes is not significant. The averaged bond length and angle values obtained from the the molecular dynamics simulations at room temperature differ from that at 0 K by less than 0.011 Å and less than one degree. With the exception of the O$^1$-C'-C$_\alpha$-H$_\alpha$ and H$^3$-N-C$_\alpha$-H$_\alpha$ torsion angles, which again correspond to rotational motions of the carboxylate and ammonium groups. Another one conspicuous discrepancy is due to the C$_\alpha$-C'-O$^1$-O$^2$ dihedral, which is normally linear in crystal and in optimized aqueous structure is close to linear. The averaged value at 300 K differ from linear by 5.8 degrees and from that at 0 K by 4.5 degrees.

[Figure 3 position]

Differences in C'-O bond lengths between crystal and water structures can be attributed to hydrogen bonding interactions. For the alanine zwitterion in water the C'-O$^1$ bond lengthens by 0.04 Å and the C'-O$^2$ bond shrinks by 0.023 Å compared to the crystalline phase. In crystal, the carboxylate group forms three hydrogen bonds; two of the hydrogen bonds involve O$^2$ and only one involves O$^1$. A schematic representation of the hydrogen bonding in the crystalline L-alanine is shown in Figure 4 (see also Figure 2 for the general structure). In aqueous solution, the number of hydrogen bonds varies. Results of molecular dynamics simulations at room temperature show that the carboxylate group can have from two to six water molecules within its first hydration shell. All of them are hydrogen-bonded to oxygen atoms of the COO$^-$ group for about 98% of the simulation time.[38] Remarkably, the C'-O$^1$ and C'-O$^2$ bond length values correlate well with the number of water molecules in the first hydration shell: as seen in Figure 5, the average values of bond lengths (computed by integration over the full molecular dynamics trajectory) demonstrate a pronounced dependence of C'-O$^1$ and C'-O$^2$ bond lengths on the number of water molecules in the vicinity of oxygen atoms. A small but significant (up to 0.05 Å) lengthening of C'-O bonds in aqueous solution compared to crystalline phase can be attributed to a larger number of hydrogen bonds in water.

[Figure 4 position]

[Figure 5 position]

The most pronounced difference between geometrical parameters in water and in crystal is related to the O$^1$-C'-C$_\alpha$-H$_\alpha$ torsion angle which indicates the COO$^-$ group orientation. Molecular dynamics simulations of alanine in aqueous solution show that all the functional sites, carboxylate, ammonium, and methyl groups, rotate with respect to $C_\alpha$-bonds. In particular, the $NH_3^+$ and $CH_3$ groups stay most of the time in the staggered conformation with respect to $H_\alpha$ and perform quick, jump-like rotations of 120 degrees. The averaged dihedral angles $H^3$-N-$C_\alpha$-$H_\alpha$ and $H^5$-$C_\beta$-$C_\alpha$-$H_\alpha$ turn out to be close to those in the crystal. The $COO^-$ group rotates more gradually, moving clockwise and counter-clockwise, with possible 60 degrees jumps.[39] In general, none of the conformations of the L-alanine molecule in water at room temperature can be considered as preferential. Similar conformational differences related to the orientation of functional sites were also established in earlier *ab initio* studies in which four[17] and nine[18] water molecules were considered to stabilize the alanine zwitterion. In crystal, strong intermolecular interactions lead to the existence of a well-defined molecular structure. The major components of the molecular motions appear to be translational in character, with a root mean-square amplitude of about 0.06 Å in all directions, and the corrections for the bond lengths should amount to less than 0.002 Å.[2] Diffusive motions are also possible; they can involve rotations of the methyl, ammonium, and carboxylate groups. However, the hydrogen-bonding interactions of the functional groups effectively increase their rotational barriers compared to those for molecules in water. Yet, the methyl group also participate in non-bonded interactions, additionally increasing barriers to rotation.[40] The rotational barriers of these groups are sufficiently high at or below room temperature.

*Intramolecular $COO^-$..$NH_3^+$ interaction.* In contrast to the crystalline phase, the $O^1$ atom of an alanine molecule in water is involved in a larger number of hydrogen bonds compared to the $O^2$ atom. On average, the $O^1$ atom is hydrogen bonded with 2.34 water molecules and the $O^2$ atom with 1.88. This partially explains the fact why the C'-$O^1$ bond is on average longer than the C'-$O^2$ bond (Table 1). A similar difference regarding the number of hydrogen bonds is also found in $NH_3$ group. The time-averaged numbers of hydrogen bonds per $H^1$, $H^2$ and $H^3$ atoms are 0.78, 0.92 and 1.0, respectively. This may be attributed to the proximity of the ammonium and carboxylate groups ($^+NH_3$...$COO^-$) and possible intramolecular hydrogen bonding interaction between them.

[Figure 6 position]

The time evolution of $H^1$-$O^1$, $H^2$-$O^2$ and N-$O^1$ distances over a course of MD simulations presented in Figure 6 indicates how often the atoms are within their van der Waals radii (2.72 Å for H-O and 3.07 for N-O). The hydrogen atom H$^1$ is mainly oriented towards the CO$_2^-$ group and is close enough to interact with the O$^1$ atom as a minor part of the bifurcated bond. This interaction must be weak because the angular geometry is unfavorable (N-H$^1$..O$^1$ < 120 degrees). The H$^2$ atom has much lower probability to be in contact with the O$^1$ atom.

The conformation in which the COOH group adopts a *cis* configuration and two intramolecular hydrogen bonds are formed between the amino group and the carbonyl oxygen is the most energetically favorable for an isolated nonionic form of L-alanine.[41] In the gas-phase intramolecular interaction dominates, but in solution and in crystal, it competes with the intermolecular hydrogen bonds and generally loses. The deformation density maps calculated as a difference between the fully self-consistent charge densities and the superposition of the atomic densities[42] are depicted in Figure 7. The planes are defined by the N-H$^1$..O$^1$ atoms. The lone-pair polarization towards the NH$^1$ atom is evident for the crystal structure (Figure 7a) and the structure in water with the COO$^-$ in the NC$_\alpha$C' plane (Figure 7b) giving a rise to a weak interaction as a minor part of the bifurcated bond. The possibility of the interaction is not obvious when the COO$^-$ is in perpendicular position with respect to the NC$_\alpha$C' plane (Figure 7c).

[Figure 7 position]

*Role of intermolecular hydrogen bonding.* An accurate analysis of an isolated L-alanine molecule showed that in the gas phase an alanine molecule exists in its neutral nonionic form (NH$_2$...COOH),[41] so that no stable conformation was found for an isolated zwitterion.[17,39] Thus, the hydrogen bonding interactions are a key factor for the existence of the stable zwitterionic structure of α-alanine in aqueous solution and in crystal.

To demonstrate that the most energetically favorable state for an alanine's zwitterion in water corresponds to the state with the maximal possible number of hydrogen bonds between the zwitterion and water molecules, we performed several additional simulations with the different number of water molecules in the first hydration shell of an alanine. For doing that, we extracted a dozen of such structures from the molecular dynamics trajectory and optimized them by searching for a minimum energy structure. Initially, the hydration level in these structures ranged from one to three water molecules per oxygen atom of the carboxylate group and from zero or one water molecule per hydrogen atom of the ammonium group. Remarkably, all the optimized structures ended up having three water molecules hydrogen bonded with the ammonium group (one per hydrogen atom) and four to six water molecules bonded to the carboxylate group (either two per each oxygen atom, or two or three on $O^2$ atom and three on $O^1$ atom). Thus, the largest number of H-bond interactions in water is energetically favorable. It should be, however, mentioned that a bifurcated hydrogen bond, with the exception of intramolecular bond described above, does not appear at any hydrogen atom of the ammonium group either in optimized structures or during molecular dynamics simulations, although *ab initio* calculations have predicted that usual and bifurcated (three-centered) hydrogen bonds have comparable energies.[43] In other zwitterion amino acids structures, the bifurcated bonds account for about 75% of the hydrogen bonds.[44]

[Table 2 position]

The geometrical parameters of the hydrogen bonds summarized in Table 2 indicate that the bond lengths are shorter and the bond angles are larger in the water compared to those in crystal. This agrees well with the qualitative relation between hydrogen bond lengths and angles: the larger the angle, the shorter the bond.[45,46] It is also notable that the crystal structure relaxation tends to change the molecular geometry in order to optimize hydrogen bond interactions, mainly through rotating $COO^-$ group, see Table 1.

## 4. Summary and Conclusions

The crystalline structure of the alanine zwitterion has been known since early works by Bernal[47] and Levy and Corey.[48] In great contrast, as far as the zwitterionic form of alanine in *aqueous solution* is concerned, its conformational properties have remained poorly understood for almost 70 years. There are many experimental techniques for obtaining information about molecules in solution, but there is at present no way of discovering the actual molecule structure. As a consequence, it is generally accepted that the structure of an alanine zwitterion obtained from the crystalline phase can also be applied to alanine in aqueous solution.

In this study we have employed first-principles computer simulations to predict the structure of L-alanine amino acid in water at room temperature. We used a density-functional linear scaling approach which allows us to study relatively large molecular system at the *ab initio* level of theory.

Using this approach we were able to treat the structure of a crystalline L-alanine and of a L-alanine solvated in 50 water molecules fully quantum mechanically.

Our *in silico* experiments revealed several noticeable differences between zwitterionic forms of alanine in water and in crystal. The largest difference can be attributed to the orientation and bond lengths of the carboxylate ($COO^-$) group. We found that in aqueous solution $C'-O^1$ and $C'-O^2$ bond lengths strongly correlate with the number of water molecules which form hydrogen bonds with the $COO^-$ group. The intramolecular $COO^-..NH_3^+$ interactions appear to be the origin of irregularity seen in the intermolecular interactions: in particular, the average numbers of hydrogen bonds per $H^{1,2,3}$ atoms of the ammonium group and per $O^{1,2}$ atoms of the carboxylate group turn out to be not equal. Furthermore, the $O^1$ atom is involved in hydrogen bonding interactions more often than the $O^2$ atom. Similarly, the $H^1$ atom of the ammonium group is more frequently hydrogen bonded than the $H^2$ and $H^3$ atoms of the same group. The observed hydrogen bond lengths are shorter and the bond angles are larger in water compared to those in crystal.

Overall, our findings strongly suggest that the structure of the alanine zwitterion in aqueous solution can differ significantly from that in crystalline phase. Therefore, care has to be taken when the structure of a L-alanine acquired from crystallographic data is applied to a L-alanine molecule in water medium.

**Acknowledgments**

This work was supported by the Academy of Finland (Center of Excellence Grant 2006-2011). The computer resources were provided by the Laboratory of Physics in Helsinki University of Technology (M-grid project). I.D. thanks Dr. Iann Gerber and Andris Gulans for stimulating discussions.

**Table 1.** Selected structural parameters of the L-alanine zwitterion (see Figures 1 and 2 for notation). The first two columns correspond to simulated structures with 50 water molecules. The third column is the optimized crystal structure and the last two columns are experimental crystalline values determined from diffraction studies.[3,2] The angles and dihedral angles are given in degrees and the bond lengths in Ångströms. The geometrical parameters for the methyl and ammonium groups are taken as averages with σ the standard deviation.

|  | water (300 K) | water (0 K) | crystal (0 K) | crystal (298 K) | crystal (23 K) |
|---|---|---|---|---|---|
| $C_\alpha$-$C_\beta$ | 1.542 | 1.532 | 1.533 | 1.523 | 1.526 |
| $C_\alpha$-N | 1.505 | 1.497 | 1.468 | 1.486 | 1.488 |
| $C_\alpha$-C' | 1.565 | 1.554 | 1.558 | 1.531 | 1.535 |
| C'-$O^1$ | 1.290 | 1.301 | 1.261 | 1.240 | 1.248 |
| C'-$O^2$ | 1.273 | 1.264 | 1.287 | 1.257 | 1.267 |
| N-$H^i$ ± σ | 1.067 ± 0.013 | 1.069 ± 0.011 | 1.075 ± 0.041 | 1.035 ± 0.011 | 1.048 ± 0.032 |
| $C_\beta$-$H^i$ ± σ | 1.122 ± 0.0 | 1.115 ± 0.001 | 1.116 ± 0.003 | 1.081 ± 0.001 | 1.096 ± 0.020 |
| <C'-$C_\alpha$-N | 110.2 | 109.5 | 111.6 | 110.1 | 110.0 |
| <C'-$C_\alpha$-$C_\beta$ | 113.2 | 114.1 | 110.5 | 111.1 | 111.1 |
| <N-$C_\alpha$-$C_\beta$ | 110.6 | 110.1 | 109.1 | 109.7 | 109.8 |
| <$O^2$-C'-$O^1$ | 125.8 | 124.7 | 125.8 | 125.6 | 125.8 |
| <$C_\alpha$-C'-$O^1$ | 117.1 | 115.8 | 120.4 | 118.4 | 118.3 |
| <$C_\alpha$-C'-$O^2$ | 116.8 | 119.2 | 113.8 | 116.0 | 115.9 |
| <$H^i$-N-$H^j$ ± σ | 107.9 ± 1.02 | 107.9 ± 3.2 | 107.7 ± 1.5 | 109.0 ± 1.4 | 109.2 ± 0.56 |
| <$H^i$-$C_\beta$-$H^j$ ± σ | 108.4 ± 0.3 | 108.7 ± 0.3 | 108.3 ± 0.2 | 108.5 ± 0.4 | 108.9 ± 0.2 |
| N-$C_\alpha$-C'-$C_\beta$ | 124.7 | -123.8 | 121.5 | -121.8 | -121.8 |
| $C_\alpha$-C'-$O^1$-$O^2$ | 174.2 | 178.7 | -179.5 | 179.8 | 179.5 |
| $O^1$-C'-$C_\alpha$-$H_\alpha$ | 98.7 | -56.7 | -142.0 | -135.4 | -135.4 |
| $H^3$-N-$C_\alpha$-$H_\alpha$ | 60.4 | 51.6 | 53.0 | 53.7 | 55.5 |
| $H^5$-$C_\beta$-$C_\alpha$-$H_\alpha$ | 58.96 | 58.9 | 59.7 | 55.5 | 55.2 |

**Table 2.** Hydrogen bonds of L-alanine zwitterion. The distances are given in Å and angles in degrees.

|  | crystal |  | water |  |
|---|---|---|---|---|
| $H^1$-$O^1$ | 1.861[a] | 1.850[b] | (N)$H_3$-$O_w$ | 1.74[c] |
| $H^2$-$O^2$ | 1.780 | 1.458 | (C)$O_2$-$H_w$ | 1.74 |
| $H^3$-$O^2$ | 1.828 | 1.876 |  |  |
| N-$O^1$ | 2.853 | 2.874 | N-$O_w$ | 2.77 |
| N-$O^2$ | 2.813 | 2.579 | (C)$O_2$-$O_w$ | 2.68 |
| N-$O^2$ | 2.832 | 2.918 |  |  |
| <N-$H^1$..$O^1$ | 160.9 | 163.1 | <N-H..$O_w$ | linear[d] |
| <N-$H^2$..$O^2$ | 168.1 | 177.5 | <O..$H_w$-$O_w$ | linear |
| <N-$H^3$..$O^2$ | 163.7 | 172.0 |  |  |

[a] Non-optimized neutron diffraction structure at 298 K.[3]

[b] Fully optimized crystal structure.

[c] The nearest neighbor distances calculated from radial distribution functions.[39]

[d] Bond angles distribution obtained after applying cone correction.[39]

**Figure 1.** Ball and stick representation of L-alanine amino acid in water. Water molecules within the first hydration shell are shown in solid. Dotted lines indicate hydrogen bonding.

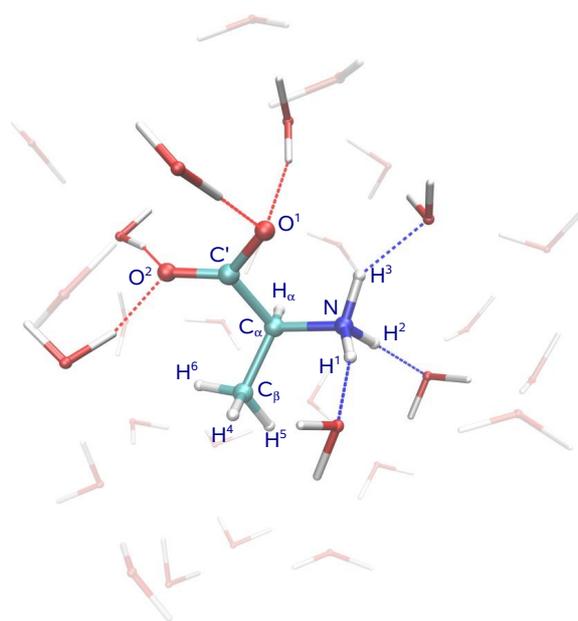

**Figure 2.** L-alanine crystalline structure. Two unit cells with species constrained to periodic boundary conditions. The system crystallizes in the $P2_12_12_1$ space group, with four zwitterionic molecules per unit cell. Adopted from Lehmann *et al.*.[3]

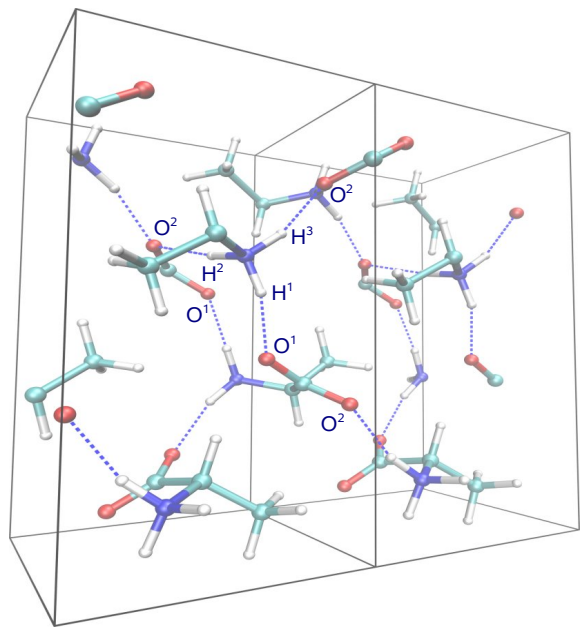

**Figure 3.** The structural differences of the L-alanine zwitterion in aqueous and crystalline structures. The calculated (blue), experimental (silver) crystalline and solvated molecule (red) structures. The structures are aligned along $C_a$-$H_a$ bond.

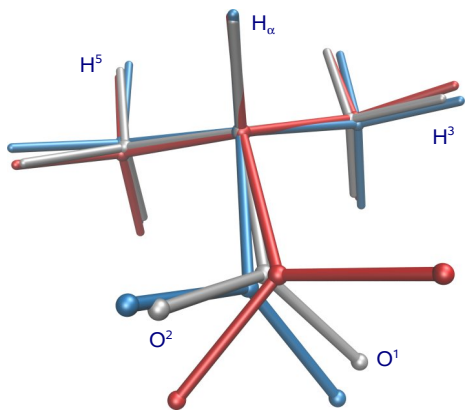

**Figure 4.** Hydrogen bonding in crystalline L-alanine.

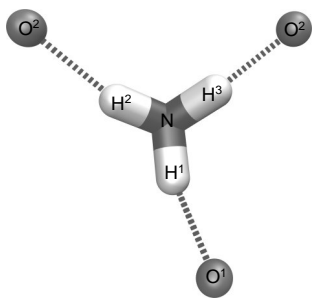

**Figure 5.** Bond lengths C'-O$^1$ and C'-O$^2$ versus number of water molecules within the hydration radius (3.06 Å for carboxylate group) around atoms O$^1$ and O$^2$ respectively. Statistical averages are computed by integrating over a full course of the molecular dynamics trajectory. The corresponding bond lengths in the experimental crystal structure[2] and in the calculated one are marked.

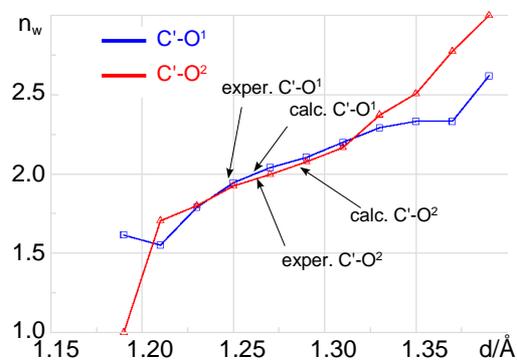

**Figure 6.** Time evolution of atomic distances during molecular dynamics simulations. H$^1$..O$^1$ distance is shown by solid line, H$^2$..O$^1$ by dashed line, and N..O$^1$ by dotted line. The running averages are taken every 500 fs.

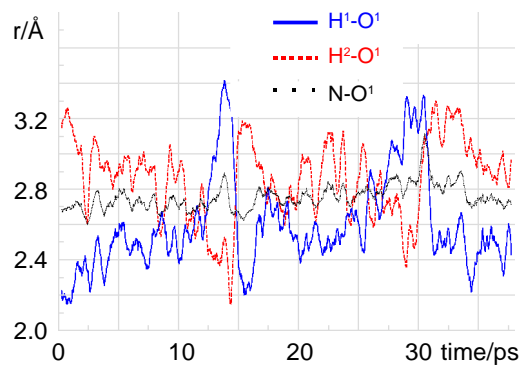

**Figure 7.** Deformation electron density maps for L-alanine molecule in aqueous solution. The contour map size is 6 × 6 Å, and the contour levels are at intervals of 0.05 e·Å$^{-3}$. The planes are selected to visualize intramolecular hydrogen bonding interactions. (a) A plane defined by H$^1$, N and O$^1$ atoms in crystalline L-alanine, (b) in aqueous L-alanine, where carboxylate is in the C'C$_\alpha$N plane, and thus O$^1$ and H$^1$ atoms are at the shortest distance, (c) in aqueous L-alanine, with the COO$^-$ perpendicular to C'C$_\alpha$N.

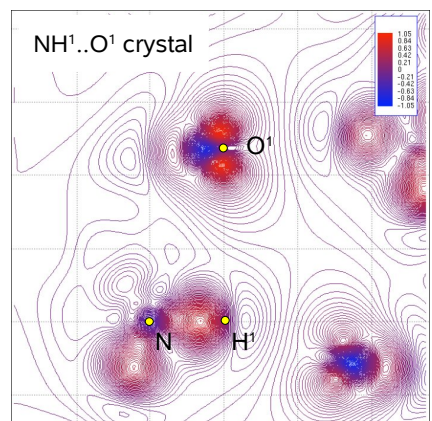

(a)

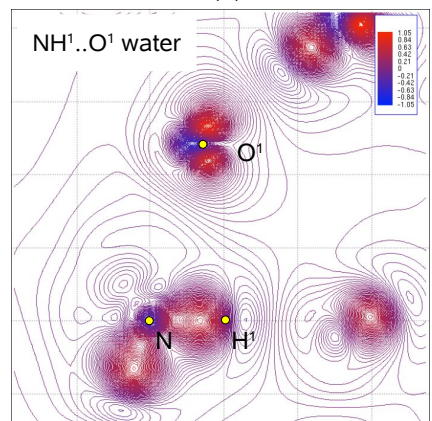

(b)

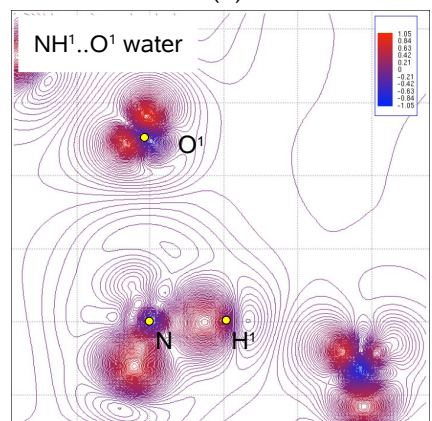

(c)